\documentclass[aps,preprint,amsmath,amssymb]{revtex4}

\usepackage{graphicx}
\begin{document}

\title{Search for electromagnetic properties of the neutrinos at the LHC}

\author{\.{I}. \c{S}ahin}
\email[]{inancsahin@karaelmas.edu.tr}
 \affiliation{Department of Physics, Zonguldak Karaelmas University, 67100 Zonguldak, Turkey}

\author{M. K\"{o}ksal}
\email[]{mkoksal@cumhuriyet.edu.tr} \affiliation{Department of
Physics, Zonguldak Karaelmas University, 67100 Zonguldak, Turkey}
\affiliation{Department of Physics, Cumhuriyet University, 58140
Sivas, Turkey}

\begin{abstract}
Exclusive production of neutrinos via photon-photon fusion provides
an excellent opportunity to probe electromagnetic properties of the
neutrinos at the LHC. We explore the potential of processes $pp\to
p\gamma\gamma p\to p\nu \bar {\nu} p$ and $pp\to p\gamma\gamma p\to
p\nu \bar {\nu}Z p$ to probe neutrino-photon and neutrino-two photon
couplings. We show that these reactions provide more than seven
orders of magnitude improvement in neutrino-two photon couplings
compared to LEP limits.
\end{abstract}

\maketitle

\section{Introduction}
Neutrinos do not interact with photons in the standard model (SM).
However, minimal extension of the SM with massive neutrinos yields
neutrino-photon and neutrino-two photon interactions through
radiative corrections
\cite{Schrock,Marciano,Lynn,Crewther,Feinberg}. Despite the fact
that minimal extension of the SM induces very small couplings, there
are several models beyond the SM predicting relatively large
neutrino-photon and neutrino-two photon couplings. Electromagnetic
properties of the neutrinos have important implications on particle
physics, astrophysics and cosmology.
Therefore probing electromagnetic structure of the neutrinos at
colliders is important for understanding the physics beyond the SM
and contributes to the studies in astrophysics and cosmology.

The Large Hadron Collider (LHC) generates high energetic
proton-proton collisions with a high luminosity.  It is commonly
believed that it will help to answer many fundamental questions in
particle physics. Recently a new phenomenon called exclusive
production was observed in the measurements of CDF collaboration
\cite{cdf1,cdf2,cdf3,cdf4,cdf5,cdf6,cdf7} and its physics potential
has being studied at the LHC
\cite{lhc1,lhc2,lhc3,avati,lhc4,royon,lhc5,lhc6,albrow,lhc7,lhc8,lhc9,lhc10,lhc11,lhc12}.
Complementary to proton-proton interactions, studies of exclusive
production might be possible and opens new field of studying very
high energy photon-photon and photon-proton interactions. Therefore
it is interesting to investigate the potential of LHC as a photon
collider to probe the electromagnetic properties of the neutrinos.

ATLAS and CMS collaborations have a program of forward physics with
extra detectors located at distances of 220m and 420m from the
interaction point \cite{royon,albrow}. Range of these 220m and 420m
detectors overlap and they can detect protons in a continuous range
of $\xi$ where $\xi$ is the proton momentum fraction loss defined by
the formula, $\xi=(|\vec{p}|-|\vec{p}^{\,\,\prime}|)/|\vec{p}|$.
Here $\vec{p}$ is the momentum of incoming proton and
$\vec{p}^{\,\,\prime}$ is the momentum of intact scattered proton.
ATLAS Forward Physics (AFP) Collaboration proposed an acceptance of
$0.0015<\xi<0.15$. There are also other scenarios with different
acceptances of the forward detectors. CMS-TOTEM forward detector
scenario spans $0.0015<\xi<0.5$ \cite{avati}. One of the well-known
application of the forward detectors is the high energy
photon-photon fusion. This reaction is produced by two quasireal
photons emitted from protons. Since the emitted quasireal photons
have a low virtuality they do not spoil the proton structure.
Therefore scattered protons are intact and forward detector
equipment allows us to detect intact scattered protons after the
collision.

The photon-photon fusion can be described by equivalent photon
approximation (EPA) \cite{budnev,Baur,lhc2}. In the framework of
EPA, emitted photons have a low virtuality  and scattered with small
angles from the beam pipe. Therefore they are almost real and the
cross section for the complete process $pp\to p\gamma\gamma p\to
pXp$ can be obtained by integrating the cross section for the
subprocess $\gamma \gamma \to X$ over the effective photon
luminosity $\frac{d L^{\gamma \gamma}}{dW}$

\begin{eqnarray}
\label{completeprocess}
 d\sigma=\int{\frac{dL^{\gamma\gamma}}{dW}}
\,d\hat {{\sigma}}_{\gamma\gamma \to X}(W)\,dW
\end{eqnarray}
where W is the invariant mass of the two photon system and the
effective photon luminosity is given by the formula
\begin{eqnarray}
\label{efflum}
\frac{dL^{\gamma\gamma}}{dW}=\int_{Q^{2}_{1,min}}^{Q^{2}_{max}}
{dQ^{2}_{1}}\int_{Q^{2}_{2,min}}^{Q^{2}_{max}}{dQ^{2}_{2}} \int_{y_{
min}}^{y_{max}} dy \frac{W}{2y} f_{1}(\frac{W^{2}}{4y}, Q^{2}_{1})
f_{2}(y,Q^{2}_{2})
\end{eqnarray}
with
\begin{eqnarray}
y_{min}=\mbox{MAX}(W^{2}/(4\xi_{max}E), \xi_{min}E),\;\;
y_{max}=\xi_{max}E,\;\;Q^{2}_{max}=2\mbox{GeV}^2
\end{eqnarray}
Here $y$ is the energy of one of the emitted photons from the
proton, $f_1$ and $f_2$ are the functions of the equivalent photon
spectra. Equivalent photon spectrum of virtuality $Q^2$ and energy
$E_\gamma$ is given by

\begin{eqnarray}
f=\frac{dN}{dE_{\gamma}dQ^{2}}
=\frac{\alpha}{\pi}\frac{1}{E_{\gamma}Q^{2}}
[(1-\frac{E_{\gamma}}{E}) (1-\frac{Q^{2}_{min}}{Q^{2}})F_{E}
+\frac{E^{2}_{\gamma}}{2E^{2}}F_{M}]
\end{eqnarray}
where
\begin{eqnarray}
Q^{2}_{min}&&=\frac{m^{2}_{p}E^{2}_{\gamma}}{E(E-E_{\gamma})},\;\;
F_{E}=\frac{4m^{2}_{p}G^{2}_{E}+Q^{2}G^{2}_{M}}
{4m^{2}_{p}+Q^{2}}\\
G^{2}_{E}&&=\frac{G^{2}_{M}}{\mu^{2}_{p}}=(1+\frac{Q^{2}}{Q^{2}_{0}})^{-4},
\;\;\; F_{M}=G^{2}_{M}, \;\;\; Q^{2}_{0}=0.71 \mbox{GeV}^{2}
\end{eqnarray}
Here E is the energy of the incoming proton beam and $m_{p}$ is the
mass of the proton. The magnetic moment of the proton is taken to be
$\mu^{2}_{p}=7.78$. $F_{E}$ and $F_{M}$ are functions of the
electric and magnetic form factors. In the above EPA formula,
electromagnetic form factors of the proton have been taken into
consideration.

In this work we investigate the potential of exclusive $pp\to
p\gamma\gamma p\to p\nu \bar {\nu} p$ and $pp\to p\gamma\gamma p\to
p\nu \bar {\nu}Z p$ reactions at the LHC to probe
$\nu\bar{\nu}\gamma$ and $\nu\bar{\nu}\gamma\gamma$ couplings. We
obtain model independent bounds on these couplings considering Dirac
neutrinos. During numerical analysis we assume that center of mass
energy of the proton-proton system is $\sqrt s=14$ TeV.

\section{Model independent analysis}
Non-standard $\nu\bar{\nu}\gamma$ interaction can be described by
the following dimension 6 effective lagrangian
\cite{Larios1,Maya,Larios2,Bell}
\begin{eqnarray}
\label{nunuphoton} {\cal
L}=\frac{1}{2}\mu_{ij}\bar{\nu}_{i}\sigma_{\mu\nu}\nu_{j}F^{\mu\nu}
\end{eqnarray}
where $\mu_{ii}$ is the magnetic moment of $\nu_i$ and $\mu_{ij}$
$(i\neq j)$ is the transition magnetic moment. In the above
effective lagrangian new physics energy scale $\Lambda$ is absorbed
in the definition of $\mu_{ii}$. The most general dimension 7
effective lagrangian describing $\nu\bar{\nu}\gamma\gamma$ coupling
is given by \cite{Nieves,Ghosh,Feinberg,Liu,Gninenko,Larios2}
\begin{eqnarray}
\label{nunuphotonphoton} {\cal
L}=\frac{1}{4\Lambda^3}\bar{\nu}_{i}\left(\alpha^{ij}_{R1} P_R+
\alpha^{ij}_{L1} P_L\right)\nu_{j}\tilde
{F}_{\mu\nu}F^{\mu\nu}+\frac{1}{4\Lambda^3}\bar{\nu}_{i}\left(\alpha^{ij}_{R2}
P_R+ \alpha^{ij}_{L2} P_L\right)\nu_{j} F_{\mu\nu}F^{\mu\nu}
\end{eqnarray}
where $P_{L(R)}=\frac{1}{2}(1\mp\gamma_5)$, $\tilde
{F}_{\mu\nu}=\frac{1}{2}\epsilon_{\mu\nu\alpha\beta}F^{\alpha\beta}$,
$\alpha^{ij}_{Lk}$ and $\alpha^{ij}_{Rk}$ are dimensionless coupling
constants.

Current experimental bounds on neutrino magnetic moment are
stringent. The most sensitive bounds from neutrino-electron
scattering experiments with reactor neutrinos are at the order of
$10^{-11}\mu_B$ \cite{Li,Wong1,Wong2,Daraktchieva}. Bounds derived
from solar neutrinos are at the same order of magnitude
\cite{Arpesella}. Bounds on magnetic moment can also be derived from
energy loss of astrophysical objects. These give about an order of
magnitude more restrictive bounds than reactor and solar neutrino
probes
\cite{Raffelt,Castellani,Catelan,Ayala,Barbieri,Lattimer,Heger}.
 Neutrino-two photon coupling has been less studied in the
literature. Current experimental bounds on this coupling are derived
from rare decay $Z\to \nu \bar{\nu} \gamma\gamma$ \cite{Larios2} and
the analysis of $\nu_\mu N\to \nu_s N$ conversion \cite{Gninenko}.
LEP data on $Z\to \nu \bar{\nu} \gamma\gamma$ decay sets an upper
bound of \cite{Larios2}
\begin{eqnarray}
\label{leplimit} \left[\frac{1 GeV}{\Lambda}\right]^6
\sum_{i,j,k}\left(|\alpha^{ij}_{Rk}|^2+|\alpha^{ij}_{Lk}|^2\right)\leq2.85\times10^{-9}
\end{eqnarray}
The analysis of the Primakoff effect on $\nu_\mu N\to \nu_s N$
conversion in the external Coulomb field of the nucleus $N$ yields
about two orders of magnitude more restrictive bound than $Z\to \nu
\bar{\nu} \gamma\gamma$ decay \cite{Gninenko}.

In the presence of the effective interactions (\ref{nunuphoton}) and
(\ref{nunuphotonphoton}), $\gamma\gamma\to \nu \bar {\nu}$
scattering is described by three tree-level diagrams. The
polarization summed amplitude square is given by the following
simple formula
\begin{eqnarray}
\label{amplitude1}
\langle|M|^2\rangle=4\sum_{i,j,m,n}\mu_{im}\mu_{mj}\mu^\ast_{in}\mu^\ast_{nj}tu+\frac{s^3}{32\Lambda^6}
\sum_{i,j,k}\left(|\alpha^{ij}_{Rk}|^2+|\alpha^{ij}_{Lk}|^2\right)
\end{eqnarray}
where s,t and u are the Mandelstam variables and we omit the mass of
neutrinos. In the above amplitude we also neglect interference terms
between interactions (\ref{nunuphoton}) and
(\ref{nunuphotonphoton}).

Neutrinos are not detected directly in the central detectors.
Instead, their presence is inferred from missing energy signal.
Therefore statistical analysis has to be performed with some care.
Any SM process with final states which are not detected by the
central detectors can not be discerned from $\gamma\gamma\to \nu
\bar {\nu}$. ATLAS and CMS have central detectors with a
pseudorapidity coverage $|\eta|<2.5$. The SM processes with final
state particles scattered with very large angles from the beam pipe
may exceed the angular range of the central detectors. Hence any SM
process with final states in the interval $|\eta|>2.5$ should be
accepted as a background for $\gamma\gamma\to \nu \bar {\nu}$. There
are also other sources of backgrounds. The one which may affect our
results is the instrumental background due to calorimeter noise. The
calorimeter noise can be effectively suppressed by imposing a cut on
the transverse energy of the jets. According to Ref.\cite{CMSTDR}
the calorimeter noise is negligible for jets with a transverse
energy greater than 40 GeV ($E_T>40$ GeV). Of course the calorimeter
noise is not the only factor which affects the jet efficiency. Based
on \cite{CMSTDR} we take into account a global efficiency of 0.6.
This is actually a prudent value for the global efficiency. We have
considered the following background processes:
\begin{eqnarray}
\label{background} &&\gamma\gamma\to \ell^-\ell^+ ;\,\,\,\,\,\,\,
\ell=e,\mu,\tau \nonumber \\ &&\gamma\gamma\to q\bar
q\,;\,\,\,\,\,\,\,\,\,\,\,\,q=u,d,s,c,b \nonumber \\
&&\gamma\gamma\to W^+W^- ;\,\,\,\,\,\,\,W^\mp\to q \bar q
^\prime,\ell \bar \nu
\end{eqnarray}
Our backgrounds can be arranged into three classes:

(1)- $\gamma\gamma\to \ell^-\ell^+$, $\gamma\gamma\to q \bar q$ and
$\gamma\gamma\to W^+W^-\to (q_1 {\bar q_1^\prime},\bar \ell_1
\nu_1)(q_2 {\bar q_2^\prime},\ell_2 \bar \nu_2)$ with $|\eta|>2.5$
for all final charged particles.

(2)-$\gamma\gamma\to q \bar q$ and $\gamma\gamma\to W^+W^- \to (q_1
{\bar q_1^\prime},\bar \ell_1 \nu_1)(q_2 {\bar q_2^\prime}, \ell_2
\bar \nu_2)$ with $|\eta|>2.5$ for final charged leptons and
$|\eta|<2.5$, $E_T<40$ GeV for final quarks, i.e., final quarks with
$E_T<40$ GeV are assumed to be missing even for $|\eta|<2.5$.

(3)-We assume that 40\% of the number of events from reactions
$\gamma\gamma\to q \bar q$ and $\gamma\gamma\to W^+W^-\to (q_1 {\bar
q_1^\prime},\bar \ell_1 \nu_1)(q_2 {\bar q_2^\prime}, \ell_2 \bar
\nu_2)$ with $|\eta|>2.5$ for final charged leptons and
$|\eta|<2.5$, $E_T>40$ GeV for final quarks is missing.

Among all the background processes the biggest contribution comes
from electron-positron production with $|\eta|>2.5$. The reason
originates from the fact that the cross section is highly peaked in
the forward and backward directions due to small mass of the
electron and $t, u = 0$ pole structure. The sum of all background
cross sections is $\sigma_{SM}=27.24$ pb for $0.0015<\xi<0.5$ and
$\sigma_{SM}=26.80$ pb for $0.0015<\xi<0.15$.

We have estimated 95\% confidence level (C.L.) bounds using
one-parameter $\chi^2$ test without a systematic error. SM
prediction about missing number of events is
$N_{SM}=0.9L_{int}\sigma_{SM}$. Here $L_{int}$ is the integrated
luminosity, 0.9 is the QED two-photon survival probability and
$\sigma_{SM}$ is the sum of all background cross sections. In order
to test SM prediction we use the following $\chi^2$ function:
\begin{eqnarray}
\label{chi2} \chi^{2}=\left(\frac{\sigma_{\gamma\gamma\to \nu
\bar{\nu}}}{\sigma_{SM} \,\, \delta_{stat}}\right)^{2}
\end{eqnarray}
where $\sigma_{\gamma\gamma\to \nu \bar {\nu}}$ is the cross section
for the process $\gamma\gamma\to \nu \bar{\nu}$ and
$\delta_{stat}=\frac{1}{\sqrt{N_{SM}}}$ is the statistical error.

In tables \ref{tab1} and \ref{tab2}, we show 95\% C.L. upper bounds
of the couplings
$\mu=\left(\sum_{i,j,m,n}\mu_{im}\mu_{mj}\mu^\ast_{in}\mu^\ast_{nj}\right)^{\frac{1}{4}}$
and
$\alpha^2=\sum_{i,j,k}\left(|\alpha^{ij}_{Rk}|^2+|\alpha^{ij}_{Lk}|^2\right)$.
As we have mentioned, diagonal elements $\mu_{ii}$ are strictly
constrained by the experiments. If we assume that $\mu_{ij}$ is
diagonal then our limits are many orders of magnitude worse than the
current experimental limits. On the other hand we see from the
tables that our bounds on $\alpha^2$ are approximately at the order
of $10^{-16}$. It is 7 orders of magnitude more restrictive than the
LEP bound. As we have mentioned, during statistical analysis we
consider the background processes given in (\ref{background}). The
main contribution is provided by the processes $\gamma\gamma\to e^-
e^+ ,\mu^- \mu^+, u \bar u$ with $|\eta|>2.5$. Of course there are
other backgrounds that we have not taken into account. But these
others are expected to give relatively small contributions.
Furthermore, even a large background does not spoil our limits
significantly. For instance, if we assume that background cross
section is 4 times larger than the sum of all backgrounds that we
have considered, our limits are spoiled only a factor of 2. They are
still at the order of $10^{-16}$.

\begin{figure}
\includegraphics[scale=1]{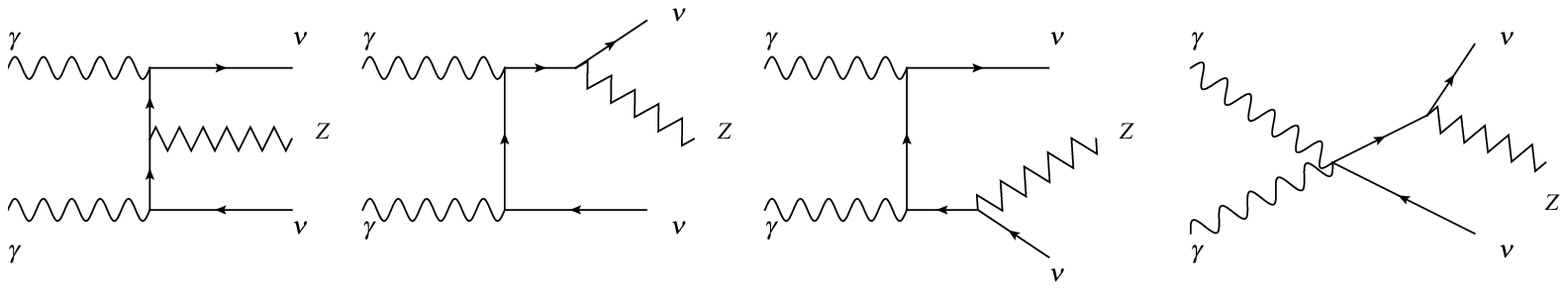}
\caption{Tree-level Feynman diagrams for the subprocess
$\gamma\gamma \to \nu \bar {\nu}Z$. The crossed diagrams are not
shown. \label{fig1}}
\end{figure}

The subprocess $\gamma\gamma \to \nu \bar {\nu}Z$ is described by 8
tree-level diagrams containing effective $\nu\bar{\nu}\gamma$ and
$\nu\bar{\nu}\gamma\gamma$ couplings (Fig.\ref{fig1}). The
analytical expression for the amplitude square is quite lengthy so
we do not present it here. But it depends on the couplings of the
form; $\sum_{i,j,k}|\alpha^{ij}_{Lk}|^2$,
 $\sum_{i,j,k}|\alpha^{ij}_{Rk}|^2$ and
$\sum_{i,j,m,n}\mu_{im}\mu_{mj}\mu^\ast_{in}\mu^\ast_{nj}$ where we
assume that $\mu_{kl}=\mu_{lk}$; $k,l=1,2,3$. $\gamma\gamma \to \nu
\bar {\nu}Z$ is absent in the SM at the tree-level. SM contribution
is originated from loop diagrams involving 5 vertices. Since the SM
contribution is very suppressed it is appropriate to set bounds on
the couplings using a Poisson distribution. The expected number of
events has been calculated considering the leptonic decay channel of
the Z boson as the signal $N=0.9L_{int}\sigma BR(Z\to \ell \bar
\ell)$, where $\ell=e^-$ or $\mu^-$. We also place a cut of
$|\eta|<2.5$ for final state $e^-$ and $\mu^-$. 95\% C.L. upper
bounds of the couplings
$\alpha_L^2=\sum_{i,j,k}|\alpha^{ij}_{Lk}|^2$,
 $\alpha_R^2=\sum_{i,j,k}|\alpha^{ij}_{Rk}|^2$ and
$\mu=\left(\sum_{i,j,m,n}\mu_{im}\mu_{mj}\mu^\ast_{in}\mu^\ast_{nj}\right)^{\frac{1}{4}}$
are presented in tables \ref{tab3} and \ref{tab4}. We observe from
the tables that the subprocess $\gamma\gamma \to \nu \bar {\nu}Z$
provides approximately an order of magnitude more restrictive bounds
on $\nu \bar \nu \gamma\gamma$ coupling with respect to
$\gamma\gamma \to \nu \bar {\nu}$. On the other hand both processes
have almost same potential to probe the coupling $\mu$.

\section{Conclusions}
Forward detector equipments allow us to study LHC as a high energy
photon collider. By use of forward detectors we can detect intact
scattered protons after the collision. Therefore deep inelastic
scattering which spoils the proton structure, can be easily
discerned from the exclusive photo-production processes. This
provides us an opportunity to probe electromagnetic properties of
the neutrinos in a very clean environment.

We show that exclusive $pp\to p\gamma\gamma p\to p\nu \bar {\nu} p$
and $pp\to p\gamma\gamma p\to p\nu \bar {\nu}Z p$ reactions at the
LHC probe neutrino-two photon couplings with a far better
sensitivity than the current limits. Former reaction improves the
sensitivity limits by up to a factor of $10^7$ and latter improves a
factor of $10^8$ with respect to LEP limits.


\newpage

\begin{table}
\caption{95\% C.L. upper bounds of the couplings $\mu$ and
$\alpha^2$ for the process $pp\to p\gamma\gamma p\to p\nu \bar {\nu}
p$. We consider various values of the integrated LHC luminosities.
Forward detector acceptance is $0.0015<\xi<0.5$. $\Lambda$ is taken
to be 1 GeV and limits of $\mu$ is given in units of Bohr magneton.
\label{tab1}}
\begin{ruledtabular}
\begin{tabular}{ccccc}
Luminosity:& 30$fb^{-1}$ &50$fb^{-1}$ &100$fb^{-1}$&200$fb^{-1}$ \\
\hline $\mu$ &2.66$\times10^{-5}$
&2.50$\times10^{-5}$ &2.29$\times10^{-5}$ &2.10$\times10^{-5}$ \\
$\alpha^2$
&2.31$\times10^{-16}$ &1.79$\times10^{-16}$ &1.27$\times10^{-16}$ &8.96$\times10^{-17}$   \\
\end{tabular}
\end{ruledtabular}
\end{table}

\begin{table}
\caption{The same as table \ref{tab1} but for $0.0015<\xi<0.15$.
\label{tab2}}
\begin{ruledtabular}
\begin{tabular}{ccccc}
Luminosity:& 30$fb^{-1}$ &50$fb^{-1}$ &100$fb^{-1}$&200$fb^{-1}$ \\
\hline $\mu$ &3.11$\times10^{-5}$
&2.91$\times10^{-5}$ &2.67$\times10^{-5}$ &2.45$\times10^{-5}$ \\
$\alpha^2$
&1.79$\times10^{-15}$ &1.39$\times10^{-15}$ &9.82$\times10^{-16}$ &6.93$\times10^{-16}$   \\
\end{tabular}
\end{ruledtabular}
\end{table}

\begin{table}
\caption{95\% C.L. upper bounds of the couplings $\mu$ and
$\alpha_{L(R)}^2$ for the process $pp\to p\gamma\gamma p\to p\nu
\bar {\nu}Z p$. We consider various values of the integrated LHC
luminosities. Forward detector acceptance is $0.0015<\xi<0.5$.
$\Lambda$ is taken to be 1 GeV and limits of $\mu$ is given in units
of Bohr magneton. \label{tab3}}
\begin{ruledtabular}
\begin{tabular}{ccccc}
Luminosity:& 30$fb^{-1}$ &50$fb^{-1}$ &100$fb^{-1}$&200$fb^{-1}$ \\
\hline $\mu$
&$2.22\times10^{-5}$ &$1.95\times10^{-5}$ &$1.64\times10^{-5}$ &$1.38\times10^{-5}$ \\
$\alpha_L^2$
&$4.65\times10^{-17}$ &$2.79\times10^{-17}$ &$1.40\times10^{-17}$  &$6.98\times10^{-18}$   \\
$\alpha_R^2$
&$4.65\times10^{-17}$ &$2.79\times10^{-17}$ &$1.40\times10^{-17}$  &$6.98\times10^{-18}$   \\
\end{tabular}
\end{ruledtabular}
\end{table}

\begin{table}
\caption{The same as table \ref{tab3} but for
$0.0015<\xi<0.15$. \label{tab4}}
\begin{ruledtabular}
\begin{tabular}{ccccc}
Luminosity:& 30$fb^{-1}$ &50$fb^{-1}$ &100$fb^{-1}$&200$fb^{-1}$ \\
\hline $\mu$
&$3.64\times10^{-5}$ &$3.20\times10^{-5}$ &$2.70\times10^{-5}$ &$2.26\times10^{-5}$ \\
$\alpha_L^2$
&$2.43\times10^{-15}$ &$1.46\times10^{-15}$ &$7.30\times10^{-16}$  &$3.65\times10^{-16}$   \\
$\alpha_R^2$
&$2.43\times10^{-15}$ &$1.46\times10^{-15}$ &$7.30\times10^{-16}$  &$3.65\times10^{-16}$   \\
\end{tabular}
\end{ruledtabular}
\end{table}

\end{document}